\documentclass[aps,prl,twocolumn,longbibliography]{revtex4-2}
\usepackage{amssymb}
\usepackage{amsbsy}
\usepackage{amsmath}
\usepackage{graphicx}
\usepackage{graphics}
\usepackage{setspace}
\usepackage{array}
\usepackage{color}
\usepackage{fontenc}
\usepackage{textcomp}
\usepackage{bm}
\usepackage{float}
\usepackage[bookmarks=false,linkcolor=blue,urlcolor=blue,colorlinks,citecolor=blue]{hyperref}

\newcommand{\darkred}[1]{{\color[rgb]{0.8,0,0}{#1}}}

\newcommand{\blue}[1]{{\color{blue}{#1}}}

\newcommand{\bsub}{\begin{subequations}}
\newcommand{\esub}{\end{subequations}}

\graphicspath{{../Figures/}}

\begin{document}

\title{Reading Altermagnetic Domains with Photon Drag}
\author{Sayed Ali Akbar Ghorashi}
\email{ghorashi@sas.upenn.edu}
\author{Andrew M. Rappe}
\affiliation{Department of Chemistry, University of Pennsylvania, Philadelphia, Pennsylvania, USA}
%\date{Version v3.3 --- 7 Oct 2026}

\begin{abstract}
Finite photon momentum allows second-order photocurrents in centrosymmetric space groups, where the conventional bulk photovoltaic effect is forbidden. Here we study the photon-drag response of altermagnets and show that its angular dependence reveals both the symmetry of the magnetic order and the N\'eel domain of the sample. For example, when a crystal rotation exchanges the two domains, the domain-even and domain-odd parts of the current appear in different angular harmonics, $4n$ and $4n+2$ for tetragonal $d$-wave altermagnets, so that two measurements rotated by $90^\circ$ on a single domain isolate the domain-odd current. We also find that these microscopic contributions exhibit opposite domain parities depending on light polarization. Under linear light, only the shift term is domain odd, while under circular light, the domain-odd channel switches to the injection, Fermi-surface, and contact principal-value terms. Mirror symmetries fix the angular phase of each harmonic and distinguish $d_{x^2-y^2}$, $d_{xy}$, and $g$-wave order. In contrast, conventional collinear antiferromagnets show no domain-odd photon-drag current, and in tetragonal ferromagnets the two parts share the same harmonics. Although the altermagnetic spin splitting does not require spin-orbit coupling (SOC), SOC is essential to observe the domain-odd charge current. Our work uncovers distinct signatures of altermagnetism in photon drag and provides a route to all optical magnetic readouts of altermagnets.
\end{abstract}
\maketitle

\blue{\emph{Introduction}}.---Altermagnets are collinear magnets in which the two spin sublattices are connected by crystal rotation rather than by translation or inversion \cite{Smejkal,McClarty,Jungwirth,altermagnetismreview, song2025altermagnets,hayami2019momentum}. As a result, they combine alternating spin splitting in the Brillouin zone with vanishing net magnetization. The resulting momentum-dependent spin splitting gives rise to a wide range of electronic phenomena, including anomalous transport, spin-polarized responses, unconventional superconductivity, and topological phases \cite{ghorashiALMTSC,ghorashiALMnonlinear,jiang2025altermagnetisminducedsurfacechern, fernandes2023topological,TSCALM2, fu2026light,lkf9-jgv6,fukaya2025superconducting,fu2026completemagnetichierarchybichromatically}. This splitting has now been resolved spectroscopically in various altermagnetic candidates \cite{Ding,ARPES1,ARPES2,ARPES3,ARPES4,ARPES5}, and nonlinear transport and optics have emerged as sensitive probes of the ordered state \cite{Fang,Usachev,GhorashiRappe,PhysRevB.111.184408, weber2024opticalexcitationspinpolarization,PhysRevB.110.024425,bzzy-ngcs,gly7-jzfl}. However, distinguishing the two N\'eel domains of an altermagnet remains challenging, since the domains are degenerate and do not carry a net moment.

Determining the domain is important because any response that is odd in the order parameter flips sign between the two domains, which means its measured value depends on how the sample’s population is distributed across domains. The standard approach is to reverse the magnetic state and compare the resulting measurements. An optical bulk probe that directly identifies the domain would eliminate this step. Nevertheless, most altermagnet candidates are centrosymmetric, and in such crystals the second-order photocurrent—the lowest-order optical response that generates a direct current—vanishes when the photon momentum is zero.
%
% Identifying the domain matters because any response that is odd in the order parameter changes sign between the two domains, so its measured value depends on the domain population of the sample. The common strategy is to reverse the magnetic state and compare the two measurements. A bulk optical probe that identifies the domain directly would remove this step. However, most altermagnet candidates are centrosymmetric, and in such crystals the second-order photocurrent, the lowest-order optical response that produces a dc current, vanishes at zero photon momentum.
%%%%--
\begin{figure}[t]
\centering
\begin{minipage}[c]{0.37\textwidth}
\centering
\includegraphics[width=\linewidth]{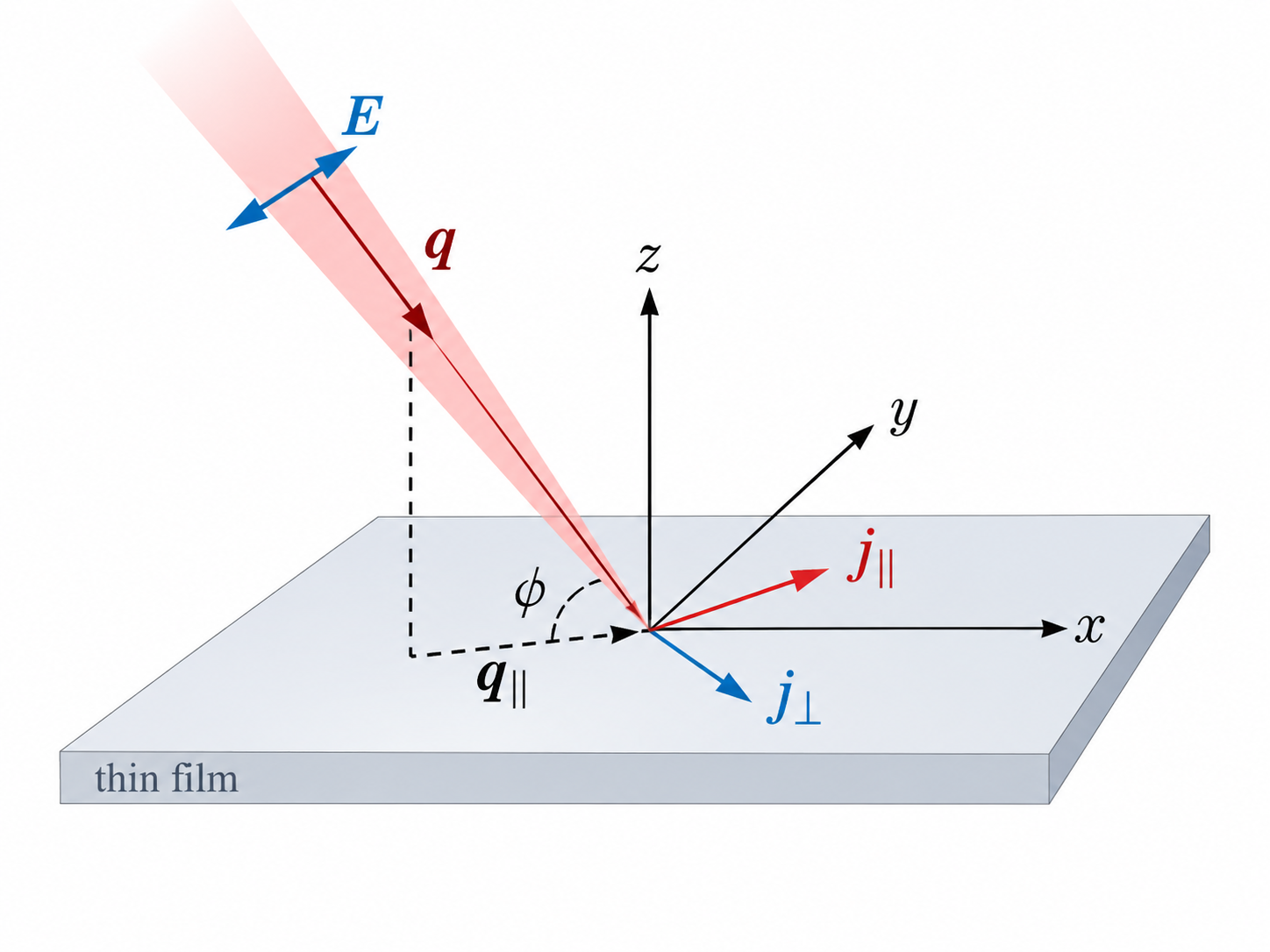}
\end{minipage}
\begin{minipage}[c]{0.5\textwidth}
\centering
{\scriptsize
\begin{ruledtabular}
\begin{tabular}{cccc}
order & magnetic-state relation & drag response\\
\hline
\\
\vspace{4pt}
$B_{1g}$ ALM & $C_4:J\to-J$ & even $4n$; odd $4n+2$; $90^\circ$ readout\\
\vspace{4pt}
tetragonal $M_z$ FM & $C_4:M_z\to M_z$ & even/odd $4n$\\
\vspace{4pt}
inversion AFM$^{\ast}$ & $P:L\to-L$ & $L$-odd $=0$\\
\vspace{4pt}
translation AFM & $\tau:L\to-L$ & $L$-odd $=0$
\end{tabular}
\end{ruledtabular}}
\vspace{2pt}
{\scriptsize $^{\ast}$All-in-plane, odd in $\bm q_\parallel$, charge-photon-drag geometry.}
\end{minipage}
\caption{\textbf{Photon-drag geometry and magnetic classes.}
(a) Oblique incidence defines the in-plane momentum $\bm q_\parallel$ and azimuth $\phi$, with current measured along or across $\bm q_\parallel$.
(b) For $B_{1g}$ altermagnets, $C_4$ interchanges the two N'eel domains, isolating domain-even from domain-odd currents across different harmonics, in contrast to conventional antiferromagnets and tetragonal ferromagnets. Here $J$, $L$, and $M_z$ represent the altermagnetic N'eel order, conventional antiferromagnetic N'eel order, and ferromagnetic magnetization, respectively.}
\label{fig:setup}
\end{figure}
%%%%%-----
%
\begin{table*}[htb!]
\caption{\textbf{Microscopic origin and angular harmonics of the photon-drag current for tetragonal $B_{1g}$ order.} Linear rows assume $\bm E\perp\bm q_\parallel$. A constant term is allowed in the $J$-even part of the linear longitudinal and circular transverse currents. For $4n+2$, $n\ge0$; for $4n$, $n\ge1$. Here \(A_0\) denotes the angle-independent (\(m=0\)) contribution to the \(J\)-even current.}
\label{tab:protocols}
\begin{ruledtabular}
\begin{tabular}{lcccc}
measurement & $J$-odd mechanism(s) & $J$-odd harmonics & $J$-even mechanism(s) & $J$-even harmonics\\
\hline
linear, $j_\perp$ & shift & $\cos(4n+2)\phi$ & inj, FS, $b\mathrm{PV}$ & $\sin4n\phi$\\
linear, $j_\parallel$ & shift & $\sin(4n+2)\phi$ & inj, FS, $b\mathrm{PV}$ & $A_0+\cos4n\phi$\\
circular, $j_\perp$ & inj, FS, $b\mathrm{PV}$ & $\sin(4n+2)\phi$ & shift & $A_0+\cos4n\phi$\\
circular, $j_\parallel$ & inj, FS, $b\mathrm{PV}$ & $\cos(4n+2)\phi$ & shift & $\sin4n\phi$
\end{tabular}
\end{ruledtabular}
\end{table*}
In this study, we demonstrate that photon drag overcomes both challenges. The finite momentum of light removes the centrosymmetry restriction on the second-order photocurrent \cite{Shi,Xie,McKay,Gassner}. We therefore examine how the angular structure of photon drag conveys information about the magnetic order. The key ingredient is the crystal rotation that defines altermagnetism. A $90^\circ$ rotation reverses the sign of the order in a $d$-wave altermagnet. In the context of photon drag, the same rotation also redirects the photon momentum, meaning that rotating the sample relative to the beam is equivalent to switching the domain. For concreteness, we focus on two-dimensional tetragonal order with the N\'eel vector lying out of plane. Nevertheless, the reasoning depends only on how the rotation acts on the domains, and we expect it to hold for other altermagnets whose domains are connected by a crystal rotation.

In the following, we demonstrate that this equivalence decomposes the photon-drag current into two components with distinct angular harmonics: one component reverses sign with the domain, and the other does not, allowing the domain to be determined from a single sample. Importantly, the microscopic mechanism whose contribution changes sign depends on the polarization: the resonant shift current for linearly polarized light, and the injection current together with the two principal-value contributions \darkred{for} circularly polarized light. Consequently, the two polarizations yield independent measurements of the same domain. We further show that the directions along which the current vanishes differ between $d_{x^2-y^2}$, $d_{xy}$, and $g$-wave order, and that neither conventional antiferromagnets nor ferromagnets reproduce the altermagnetic pattern. Our results are summarized in Fig.~\ref{fig:setup}.

\blue{\emph{Photon-drag geometry}}.---We start by defining the optical geometry, shown in Fig.~\ref{fig:setup}(a). The light incident at an oblique angle carries a full photon momentum $\bm q$ with in-plane component $\bm q_\parallel=q_\parallel(\cos\phi,\sin\phi)$, where $q_\parallel=|\bm q_\parallel|=q\sin\theta$, $q=|\bm q|$, and $\theta$ is measured from $\hat z$; the azimuthal angle $\phi$ is measured relative to a crystal axis. We denote $\hat q_\parallel=(\cos\phi,\sin\phi)$ and $\hat e_\perp=(-\sin\phi,\cos\phi)$, take a linear polarization $\hat e_\beta=\cos\beta\,\hat q_\parallel+\sin\beta\,\hat e_\perp$ at a fixed angle $\beta$ relative to $\bm q_\parallel$, and measure the current along $u_\parallel=\hat q_\parallel$ or across $u_\perp=\hat e_\perp$. The measured current is written as
\begin{equation}
 j_{\nu p}(\phi;J)=u_\nu^a\,
 \sigma_p^{a;bc}(\bm q_\parallel,\omega;J)P^p_{bc}|E|^2,
 \qquad p=L,C ,
\label{eq:geometry}
\end{equation}
where $\sigma_p^{a;bc}$ is the second-order conductivity with current index $a$ and optical indices $b,c$. For linear light $P^L_{bc}=e_{\beta,b}e_{\beta,c}$; for circularly polarized light of helicity $h=\pm1$, $P^C_{bc}=-ih\epsilon_{bc}/2$, where $\epsilon_{bc}$ is the in-plane antisymmetric tensor, $\epsilon_{xy}=+1$. We refer to the four combinations of $\nu=\parallel,\perp$ and $p=L,C$ collectively as $j_X$. We focus on two-dimensional systems with the Néel vector oriented along $\hat z$ with all three tensor indices lying in the plane. In this geometry $C_{2z}$ enforces $\sigma^{a;bc}(-\bm q_\parallel)=-\sigma^{a;bc}(\bm q_\parallel)$, eliminating the second-order $\bm q_\parallel=0$ response and so the response arises entirely from photon drag.
\begin{table*}[tb!]
\caption{\textbf{Rotational and mirror fingerprints of tetragonal orders in the transverse linear current.} Nulls are symmetry-enforced zeros of the total transverse linear current. The rotation fixes the harmonic order and the vertical mirrors fix the sine or cosine phase.}
\label{tab:representations}
\begin{ruledtabular}
\begin{tabular}{lccccc}
order & rotation of domain & $J$-even harmonics & $J$-odd harmonics & total-current nulls & separation\\
\hline
$B_{1g}$ $d$ wave & $C_4:J\to-J$ & $\sin4n\phi$ & $\cos(4n+2)\phi$ & diagonals & $90^\circ$\\
$B_{2g}$ $d$ wave & $C_4:J\to-J$ & $\sin4n\phi$ & $\sin(4n+2)\phi$ & axes & $90^\circ$\\
$A_{2g}$ $g$ wave & $C_4:J\to J$ & $\sin4n\phi$ & $\sin4n\phi$ & axes, diagonals & none in plane\\
\end{tabular}
\end{ruledtabular}
\end{table*}

\blue{\emph{Domain parity and angular harmonics}}.--- We now show how crystal rotation constrains the angular dependence of $j_X$. Let $J=\pm|J|$ label the two N\'eel domains, and let an $N$-fold proper crystal rotation act on the Hamiltonian as
\begin{equation}
 U_NH(\bm k;J)U_N^{-1}=H(R_N\bm k;\zeta J),\qquad \zeta=\pm1,
\label{eq:domain}
\end{equation}
where $U_N$ acts on the Bloch states, $R_N$ rotates the in-plane momentum by $2\pi/N$, and $\zeta=+1$ ($-1$) when the rotation preserves (exchanges) the domain. A rotation that exchanges the two domains can occur only for even $N$, since applying it $N$ times must return the system to the original domain; for crystallographic rotations this gives $N=2,4,6$. Rotating the full optical geometry by $2\pi/N$ rotates $\bm q_\parallel$, the polarization, and the analyzer together, while $J\to\zeta J$. It follows that $j_X(\phi+2\pi/N;J)=j_X(\phi;\zeta J)$. We separate the current into domain-even ($\eta=+1$) and domain-odd ($\eta=-1$) parts, $j^{(\eta)}_X(\phi;J)=[j_X(\phi;J)+\eta\, j_X(\phi;-J)]/2$. Each part then satisfies
\begin{equation}
\begin{aligned}
 j_X^{(\eta)}\!\left(\phi+\frac{2\pi}{N};J\right)
 &=\eta^{(1-\zeta)/2}j_X^{(\eta)}(\phi;J),\\[-1pt]
 c_m\neq0\quad&\Longrightarrow\quad
 e^{i2\pi m/N}=\eta^{(1-\zeta)/2},
\end{aligned}
\label{eq:harmonic}
\end{equation}
where $c_m$ is the coefficient in $j_X^{(\eta)}=\sum_m c_m e^{im\phi}$. If the rotation preserves the domain, both parts contain only harmonics $m=Nn$. If it exchanges the domains, the even part remains at $m=Nn$, while the odd part appears at $m=Nn+N/2$. The two domain parities are therefore carried by different angular harmonics. This result follows only from the rotational covariance of Eq.~\eqref{eq:geometry}; it holds for any fixed $\beta$ and for each microscopic mechanism. 

\blue{\emph{Domain readout}}.--- In a $B_{1g}$ altermagnet $C_4$ exchanges the two domains, $\zeta=-1$, and Eq.~\eqref{eq:harmonic} yields both parts from a pair of azimuths on the same domain,
\begin{equation}
\begin{aligned}
 j_{J-}(\phi)&=\frac{j(\phi;J)-j(\phi+\pi/2;J)}{2},\\
 j_{J+}(\phi)&=\frac{j(\phi;J)+j(\phi+\pi/2;J)}{2},
\end{aligned}
\label{eq:readout}
\end{equation}
where $j_{J-}$ and $j_{J+}$ denote the domain-odd ($\eta=-1$) and domain-even ($\eta=+1$) parts. Since $j_{J-}$ changes sign between the domains, its sign identifies the domain once calibrated. If the illuminated region contains fractions $p$ and $1-p$ of the two domains, the extracted signal is reduced by the factor $2p-1$, and unless they are exactly equal, the response can be detected in multi-domain systems. We emphasize that the domain-even part is not simply a nonmagnetic background, but it contains altermagnetic contributions that are even in $J$. The alignment accuracy required for the $90^\circ$ rotation is estimated in the Supplemental Material (SM) \cite{SM}.

\blue{\emph{Comparison with conventional magnets}}.---To confirm that this pattern is specific to altermagnets, we compare it with conventional collinear magnets, as summarized in Fig.~\ref{fig:setup}(b). First, consider an antiferromagnet where two magnetic sublattices are related by a lattice translation. Under this translation, the optical fields at $\bm q_\parallel$ and $-\bm q_\parallel$ acquire opposite phases, which cancel in the dc product $E_{\bm q_\parallel}E_{-\bm q_\parallel}$. Since the uniform current is unchanged by a lattice translation, the two domains produce the same photon-drag current, and the domain-odd response vanishes. Next, consider an antiferromagnet in which inversion exchanges the two magnetic sublattices. In the plane, inversion sends $\phi$ to $\phi+\pi$ and swaps the domains, whereas the photon-drag current—being odd in $\bm q_\parallel$—is itself $\pi$-periodic when the domain is held fixed. Taken together, these two constraints require the current to be even under domain exchange, while neither constraint on its own is sufficient. By contrast, a tetragonal ferromagnet with out-of-plane magnetization $M_z$ can support a drag current that changes sign with $M_z$, because both $C_{2z}$ and $C_4$ leave the magnetization unchanged. In this case $\zeta=+1$ in Eq.~\eqref{eq:harmonic}, so the even and odd responses occupy the same $4n$ harmonics and cannot be separated by rotation. Therefore, the separation relies on the fact that a rotation exchanges the two magnetic sublattices, a characteristic of altermagnets; the corresponding derivations are given in the SM.

\blue{\emph{Microscopic mechanisms}}.--- Symmetry selection determines which harmonics are allowed, but leaves their microscopic origins unspecified. Unpacking these processes requires resolving the finite-$q_\parallel$ response into four mechanism-resolved contributions. Setting the initial and final states to $i=(m,\bm k-\bm q_\parallel/2)$ and $f=(n,\bm k+\bm q_\parallel/2)$, the integrands take the form
\begin{equation}
\begin{aligned}
 {\cal K}_{\rm sh}^a&=F_{fi}|M_{fi}|^2R_{fi}^a\,\delta(\Delta_{fi}-\hbar\omega),\\
 {\cal K}_{\rm inj}^a&=\tau_{\rm inj}F_{fi}|M_{fi}|^2(v_f^a-v_i^a)\,\delta(\Delta_{fi}-\hbar\omega),\\
 {\cal K}_{\rm FS}^a&=(\partial_{k_a}F_{fi})|M_{fi}|^2\,{\cal P}\frac1{\Delta_{fi}-\hbar\omega},\\
 {\cal K}_{b{\rm PV}}^a&=2F_{fi}\operatorname{Re}(M_{fi}^*B_{fi}^a)\,{\cal P}\frac1{\Delta_{fi}-\hbar\omega},
\end{aligned}
\label{eq:kernels}
\end{equation}
Here, $F_{fi}=f_i-f_f$ is the occupation difference, $\Delta_{fi}=\epsilon_f-\epsilon_i$ the transition energy, $M_{fi}$ the finite-$q_\parallel$ optical matrix element, $R^a_{fi}$ the gauge-invariant shift vector, $v^a_{f,i}$ the band velocities at displaced momenta, \(\tau_{\rm inj}=\hbar/\Gamma_0\) is the injection relaxation time, $B^a_{fi}$ the contact vertex originating from the field dependence of the current operator, and $\mathcal P$ the principal value; explicit prefactors and full derivations are provided in the SM. The shift and injection components are strictly resonant, whereas the Fermi-surface (FS) and contact principal-value ($b\mathrm{PV}$) contributions are dispersive. The $b\mathrm{PV}$ term arises because two frequency orderings of the contact vertex combine cleanly into resonant and principal-value parts. While Ref.~\cite{Xie} absorbs the resonant piece into the shift current and drops the principal-value piece, we retain it here, noting that it feeds into the same harmonics as the FS term. Keeping all four terms lets us track which microscopic process drives each symmetry signature. The classification below holds regardless of how one groups the dispersive principal-value parts.
\begin{figure*}[htb!]
\includegraphics[width=0.95\textwidth]{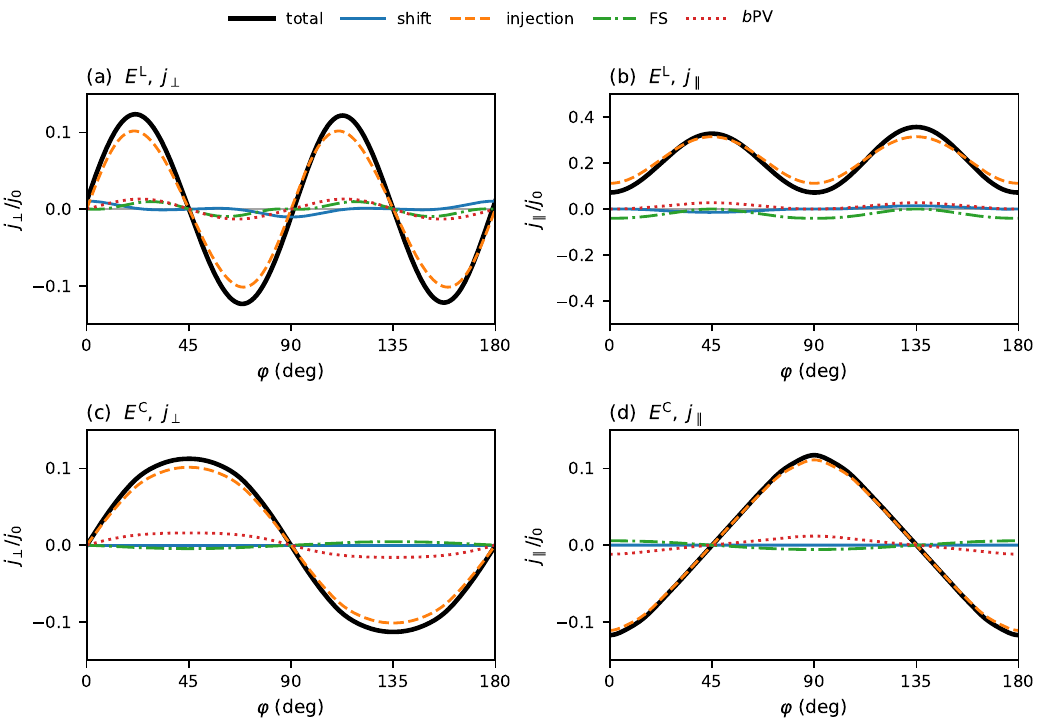}
\caption{\textbf{Photon-drag currents of the $B_{1g}$ model of Eq.~\eqref{eq:model}.} (a,b) Linear light with the current measured across and along $\bm q_\parallel$. (c,d) Circular light with the same two projections. Black solid lines show the sum of the four contributions; blue solid, orange dashed, green dash-dotted, and red dotted lines show the shift, injection, FS, and $b\mathrm{PV}$ contributions. Parameters are $J=t$, $\lambda=0.2t$, $\mu=0.2t$, $\hbar\omega=0.8t$, $k_BT=0.05t$, $q_\parallel a=0.2$, optical width $\Gamma=0.06t$, and dc width $\Gamma_0=0.04t$, corresponding to $\tau_{\rm inj}=\hbar/\Gamma_0$. Currents are normalized by $j_0=e^3a|E|^2/(\hbar t)$, with $|E|$ the positive-frequency field amplitude}.
\label{fig:atlas}
\end{figure*}
$C_{2z}$ reverses $\bm q_\parallel$, while time reversal exchanges the two N\'eel domains; together they determine the domain parity of each term. Under linearly polarized light, the shift current is odd in $J$, whereas the injection, FS, and $b\mathrm{PV}$ contributions are even. Moving to circular polarization completely flips these assignments. This swap traces back to an explicit factor of $i$ within the shift conductivity, forcing the linear shift current to pick up the imaginary part of its kernel and the circular shift current to take the real part—behavior precisely opposite to the other three terms. Combining these parities with Eq. \eqref{eq:harmonic} and the mirror symmetries yields the fingerprints of $B_{1g}$ order in Table \ref{tab:protocols}. For each polarization and detection direction, the table specifies which mechanism changes sign with the domain, which harmonics it occupies, and whether it appears as a sine or a cosine; the mirror phases assume $\beta=0$ or $\pi/2$, and the linear results below use $\beta=\pi/2$.
\emph{$d$- versus $g$-wave order}.---As outlined in Table \ref{tab:representations}, the specific magnetic order is pinned down by the combination of harmonic channels, phase shifts, and symmetry nodes where the current vanishes. While the $B_{1g}$ ($d_{x^2-y^2}$) and $B_{2g}$ ($d_{xy}$) representations both exchange domains under $C_4$, their mirror eigenvalues are inverted—causing the transverse linear response to drop to zero along the diagonal directions for $B_{1g}$ and along the principal crystal axes for $B_{2g}$. By contrast, the tetragonal $A_{2g}$ $g$-wave state follows a different symmetry requirement, possessing a form factor of:$ g_g(\bm k)=\sin k_x\sin k_y(\cos k_x-\cos k_y),$
where $g_g$ modifies the exchange coupling through $Jg_g(\bm k)\sigma_z$. Since $C_4$ symmetry leaves this term intact, rotating the system does not flip the domain. The domain-odd current persists, but it resides in the same $4n$ harmonics as the domain-even piece, and the transverse linear current goes to zero along the axes as well as the diagonals. Consequently, isolating the two contributions requires explicitly switching the domain orientation.

The domain-readout scheme formulated in Eq.~\darkred{\eqref{eq:harmonic}} extends naturally beyond fourfold rotations. When $C_6$ symmetry exchanges the domains, domain-even and domain-odd currents occupy the $6n$ and $6n+3$ harmonics, respectively. However, in a strictly two-dimensional, in-plane setup, the inherent $\pi$ periodicity eliminates these odd harmonics, rendering the $C_6$ domain-odd signal unobservable in this geometry. This limitation stems from the two-dimensional constraint rather than the core readout mechanism. In three-dimensional hexagonal altermagnets like MnTe and CrSb, the rotational symmetry logic holds, though the experimental setup must probe symmetry-allowed out-of-plane momenta or tensor elements. Likewise, if domains are mapped into each other by a horizontal mirror, an in-plane signal at $q_z=0$ is strictly domain even, whereas a 3D geometry can isolate the domain-odd contribution. A systematic treatment of these alternative geometries will be addressed elsewhere.

\emph{Lattice model}.--- Now, we demonstrate these readout recipes in a minimal two-band model of a $d_{x^2-y^2}$-wave ($B_{1g}$) altermagnet,
\begin{equation}
\begin{aligned}
H(\bm k;J)={}&t(\cos k_x+\cos k_y)\sigma_0
-\lambda\sin k_y\sigma_x+\lambda\sin k_x\sigma_y\\
&+J(\cos k_x-\cos k_y)\sigma_z,
\end{aligned}
\label{eq:model}
\end{equation}
where $t$ denotes the nearest-neighbor hopping amplitude, $\lambda$ the Rashba spin-orbit coupling strength, $J$ the altermagnetic exchange, and $\sigma_{0,x,y,z}$ denote the Pauli matrices in spin, with the lattice constant set to unity. Although the Rashba term breaks spatial inversion, $C_{2z}$ symmetry continues to enforce $\sigma^{a;bc}(-\bm q_\parallel)=-\sigma^{a;bc}(\bm q_\parallel)$ for in-plane tensor components, making the response of Eq. \eqref{eq:model} purely photon drag. Aligning the N'eel vector in-plane would break $C_{2z}$ and permit a $q_\parallel=0$ photocurrent \cite{Ezawa}; we therefore keep the N'eel vector oriented out of plane. Two exact identities of Eq. \eqref{eq:model} clarify how the magnetic order enters the response. First, for $\bm d=(-\lambda\sin k_y,\lambda\sin k_x,J(\cos k_x-\cos k_y))$, the triple product simplifies directly to the magnetic form factor,
\begin{equation}
 \bm d\cdot(\partial_{k_x}\bm d\times\partial_{k_y}\bm d)=-\lambda^2J(\cos k_x-\cos k_y),
\label{eq:berry}
\end{equation}
The Berry curvature therefore changes sign along the $k_x=\pm k_y$ nodal lines, and every domain-odd kernel contains an explicit prefactor of $\lambda^2J$—though this prefactor does not imply a trivial scaling of the total current with $\lambda$ (see SM for details). Second, in the absence of SOC ($\lambda=0$), the unitary matrix $\sigma_x$ relates $H(\bm k;J)$ to $H(\bm k;-J)$ alongside all charge vertices, ensuring the two domains produce indistinguishable charge currents. Thus, while altermagnetic band splitting survives at $\lambda=0$, an observable domain-odd charge current relies on finite SOC. We couple light via straight-bond Peierls phases, accounting for finite-$q$ effects in both the endpoint momenta and current vertices.
Figure~\ref{fig:atlas} details the response for linear and circular polarizations, decomposed into components along and across the photon momentum.
For linear polarization, the domain-odd contribution arises exclusively from the shift current, whereas for circular light it stems from the injection, FS, and $b\mathrm{PV}$ channels; all four configurations display angular phases consistent with Table~\ref{tab:protocols}. The relative magnitude of the domain-even and domain-odd components varies across the four measurement channels, but the quarter-turn subtraction in Eq.~\eqref{eq:readout} isolates the domain-odd current independently of their relative amplitudes. This robust isolation occurs because the separation is rooted in symmetry rather than signal magnitude.

\emph{Discussion}.---Several comments regarding the scope and experimental implementation are in order. First, our analysis determines the crystal representation of the magnetic order rather than its continuum angular rank, which is the property targeted by multiphoton and rank-selective optical techniques \cite{GhorashiRappe,Tagani}. Second, a symmetry-enforced zero is primarily useful for ruling out candidate orders, as a symmetry-allowed current can remain small due to domain averaging, weak matrix elements, or accidental cancellation. Furthermore, any bulk extrinsic current that respects the same magnetic point group will occupy the same harmonics as the intrinsic shift current; reversing $\bm q_\parallel$ verifies momentum parity but does not distinguish the underlying microscopic mechanism. Third, our calculations assume an in-plane optical field. Oblique $p$-polarized incidence introduces an $E_z$ component, where mixed tensor elements like $\sigma^{y;xz}$ follow distinct selection rules and must be either eliminated or separated experimentally. Finally, the in-plane photon momentum assumed here, $q_\parallel a=0.2$, serves as a finite-$q_\parallel$ benchmark: while a free-space photon at $\hbar\omega=0.8$ eV yields $qa\simeq1.6\times10^{-3}$ for $a=4$ \AA{}, so $q_\parallel a=qa\sin\theta$, polaritonic, plasmonic, or guided-wave platforms supply much larger in-plane momentum \cite{Shi,Xiong}, making them natural testbeds for these measurements. Beyond establishing the angular signatures of altermagnetism in finite-momentum photocurrents, these results provide a strategy for reading N'eel domains without magnetic field switching.

\blue{\emph{Acknowledgment}}.---This work was supported by the U.S. Department of Energy, Office of Science, Basic Energy Sciences, under Award No. DE-SC0024942. S.A.A.G acknowledges the hospitality of the Aspen Center for Physics, which is supported by National Science Foundation grant PHY-2210452 and Kavli Institute for Theoretical Physics (KITP) supported by NSF PHY-2309135.
%\clearpage
\bibliography{draft.bib}

\clearpage
\onecolumngrid
\begin{center}
{\large\bf Supplemental Material}\\[2pt]
\end{center}
\setcounter{equation}{0}
\renewcommand{\theequation}{S\arabic{equation}}
\setcounter{figure}{0}
\renewcommand{\thefigure}{S\arabic{figure}}

\section*{S1. Harmonic rule}
In this section, we derive Eq.~\eqref{eq:harmonic} of the main text and its generalization to an in-plane point group. Let $R_N$ be a proper in-plane rotation by $\theta_N=2\pi/N$ that acts on the Hamiltonian as
\begin{equation}
 U_NH(\bm k;J)U_N^{-1}=H(R_N\bm k;\zeta J),\qquad \zeta=\pm1.
\label{eq:smrot}
\end{equation}
Tensor covariance of the finite-momentum conductivity then gives
\begin{equation}
 \sigma^{a;bc}(R_N\bm q_\parallel;J)
 =(R_N)_{aa'}(R_N)_{bb'}(R_N)_{cc'}
 \sigma^{a';b'c'}(\bm q_\parallel;\zeta J).
\label{eq:smtensorcov}
\end{equation}
For fixed $\beta$ the polarization and both analyzers rotate with the photon momentum, $\hat e_\beta(\phi+\theta_N)=R_N\hat e_\beta(\phi)$ and $u_\nu(\phi+\theta_N)=R_Nu_\nu(\phi)$, while the circular dyad is invariant, $R_N\epsilon R_N^{T}=\epsilon$. Contracting Eq.~\eqref{eq:smtensorcov} with the rotated dyad and analyzer removes all rotation matrices, and we obtain
\begin{equation}
 j_X(\phi+\theta_N;J)=j_X(\phi;\zeta J)
\label{eq:smcov}
\end{equation}
for each of the four currents $X=(\nu,p)$. With the domain-even and domain-odd parts
\begin{equation}
 j_X^{(\eta)}(\phi;J)=\tfrac12\left[j_X(\phi;J)+\eta j_X(\phi;-J)\right],\qquad \eta=\pm1,
\label{eq:smparity}
\end{equation}
which satisfy $j_X^{(\eta)}(\phi;-J)=\eta j_X^{(\eta)}(\phi;J)$, Eq.~\eqref{eq:smcov} becomes $j_X^{(\eta)}(\phi+2\pi/N;J)=\eta^{(1-\zeta)/2}j_X^{(\eta)}(\phi;J)$. Expanding in $e^{im\phi}$ and using the linear independence of the harmonics yields
\begin{equation}
 c_{m,\eta}\left[e^{i2\pi m/N}-\eta^{(1-\zeta)/2}\right]=0
 \quad\Longrightarrow\quad
 e^{i2\pi m/N}=\eta^{(1-\zeta)/2}.
\label{eq:smmaster}
\end{equation}
For $\zeta=+1$ both parts have $m=Nn$. For $\zeta=-1$ the even part has $m=Nn$ and the odd part $m=Nn+N/2$; this requires even $N$, since $C_N^N$ returns the Hamiltonian to the original domain and hence $\zeta^N=1$. For $C_4:J\to-J$ one has $j(\phi+\pi/2;J)=j(\phi;-J)$, which gives Eq.~\eqref{eq:readout}. More generally, for an in-plane point group $G$ with domain character $\chi_J(g)$, the domain-odd part follows from measurements at the transformed azimuths $\phi_g$ as
\begin{equation}
 j_{J-}^{p,\nu}(\phi;J)=\frac1{|G|}\sum_{g\in G}\chi_J(g)w_{p,\nu}(g)j_{p,\nu}(\phi_g;J),
\label{eq:smgroupproj}
\end{equation}
where $w_{L,\perp}=\det R_g$, $w_{L,\parallel}=1$, $w_{C,\perp}=1$, and $w_{C,\parallel}=\det R_g$. If every in-plane operation preserves the domain, $\chi_J$ is trivial and no combination of such measurements separates the two parts.

\section*{S2. Conventional collinear magnets}
In this section, we show that conventional collinear magnets do not reproduce the altermagnetic pattern. We consider the uniform bulk dc charge current driven by optical fields at $\pm\bm q_\parallel$, with all three tensor indices in the plane.

\emph{Translation-compensated antiferromagnet}.---If a lattice translation $\bm\tau$ exchanges the two N\'eel states, the fields transform as $E_{\bm q_\parallel}\to e^{i\bm q_\parallel\cdot\bm\tau}E_{\bm q_\parallel}$ and $E_{-\bm q_\parallel}\to e^{-i\bm q_\parallel\cdot\bm\tau}E_{-\bm q_\parallel}$, so the phases cancel in the dc bilinear $E_{\bm q_\parallel}E_{-\bm q_\parallel}$. Since the uniform current is invariant under translation,
\begin{equation}
j(\bm q_\parallel;L)=j(\bm q_\parallel;-L),
\label{eq:smtransafm}
\end{equation}
and the domain-odd part vanishes. This argument applies to the uniform dc current only; probes at finite output wave vector and sublattice-resolved probes are not constrained in the same way.

\emph{Inversion-compensated antiferromagnet}.---If inversion exchanges the two N\'eel states, then, since inversion maps $\phi$ to $\phi+\pi$ in the plane, the current is a polar vector, and the quadratic dyad is inversion even,
\begin{equation}
 j_X(\phi+\pi;L)=j_X(\phi;-L).
\label{eq:sminv1}
\end{equation}
Independently, we isolate the drag part by its odd momentum parity, $\sigma_{\rm drag}^{a;bc}(-\bm q_\parallel;L)=-\sigma_{\rm drag}^{a;bc}(\bm q_\parallel;L)$, assuming that any \darkred{$q_\parallel$-even} electric-dipole contribution is absent or has been subtracted. Under $\phi\to\phi+\pi$ the analyzer reverses; for linear light both field vectors also reverse, while the circular dyad is unchanged. Together with the momentum oddness, the projected current is therefore $\pi$ periodic at fixed $L$,
\begin{equation}
 j_X(\phi+\pi;L)=j_X(\phi;L),
\label{eq:sminv2}
\end{equation}
and combining Eqs.~\eqref{eq:sminv1} and \eqref{eq:sminv2} gives
\begin{equation}
 j_X(\phi;-L)=j_X(\phi;L).
\label{eq:sminv3}
\end{equation}
Notice that momentum oddness alone does not constrain the domain parity; the conclusion requires both relations. Mixed-$z$ tensor components and other N\'eel orientations are not covered by this argument.

\emph{Tetragonal ferromagnet}.---For out-of-plane magnetization $M_z$, both $C_{2z}$ and $C_4$ preserve the magnetic state, so $\pi$ periodicity never relates $+M_z$ to $-M_z$ and a magnetization-odd drag current is allowed. However, $\zeta=+1$ in Eq.~\eqref{eq:smmaster}, so both parts have $m=4n$ and no quarter-turn subtraction separates them.

Finally, for incoherent domain fractions $p$ and $1-p$ in the illuminated region, $j_{\rm meas}=j_{J+}+(2p-1)j_{J-}$. The subtraction removes the domain-even part but cannot restore a domain-odd signal lost to balanced populations.

\section*{S3. Mirror phases and tetragonal fingerprints}
Rotations fix the allowed harmonics, and mirrors fix their sine or cosine phase. Under a vertical reflection the transverse analyzer acquires $\det R=-1$ relative to the longitudinal one, a real linear dyad with $\beta=0$ or $\pi/2$ maps onto itself, and the antisymmetric circular dyad acquires an additional pseudoscalar sign. Collecting these factors, a mirror $M$ with domain character $\chi_J(M)$ acts as
\begin{equation}
 j^{(\eta)}_{p,\nu}(\phi_M)=s_p\,a_\nu\,\eta^{[1-\chi_J(M)]/2}\,j^{(\eta)}_{p,\nu}(\phi),
 \qquad s_L=+1,\ s_C=-1,\ a_\parallel=+1,\ a_\perp=-1 ,
\label{eq:smmirror}
\end{equation}
where $\phi_M$ is the mirror image of $\phi$. A current is even about the mirror-invariant direction when the prefactor is $+1$, and it vanishes along that direction when the prefactor is $-1$. Applying the $B_{1g}$ characters, we obtain
\begin{align}
 j_{\perp,{\rm sh},L}&=\sum_{n\ge0}A_{4n+2}\cos[(4n+2)\phi],&
 j_{\parallel,{\rm sh},L}&=\sum_{n\ge0}B_{4n+2}\sin[(4n+2)\phi],\\
 j_{\perp,Y,L}&=\sum_{n\ge1}B^Y_{4n}\sin4n\phi,&
 j_{\parallel,Y,L}&=A_0^Y+\sum_{n\ge1}A^Y_{4n}\cos4n\phi,\\
 j_{\perp,{\rm sh},C}&=A_0^{\rm sh}+\sum_{n\ge1}A^{\rm sh}_{4n}\cos4n\phi,&
 j_{\parallel,{\rm sh},C}&=\sum_{n\ge1}B^{\rm sh}_{4n}\sin4n\phi,\\
 j_{\perp,Y,C}&=\sum_{n\ge0}B^Y_{4n+2}\sin[(4n+2)\phi],&
 j_{\parallel,Y,C}&=\sum_{n\ge0}A^Y_{4n+2}\cos[(4n+2)\phi],
\end{align}
with $Y\in\{{\rm inj,FS},b{\rm PV}\}$, which reproduces Table~\ref{tab:protocols}. Symmetry fixes these harmonics and phases, but not the coefficients $A_m$ and $B_m$, which are determined by the finite-$q_\parallel$ kernels.

For an out-of-plane exchange term $Jg_\Gamma(\bm k)\sigma_z$, the axial $\sigma_z$ is odd under every vertical mirror, so $\chi_J=\chi_\Gamma\chi_{\sigma_z}$ with $\chi_{\sigma_z}(C_4)=+1$ and $\chi_{\sigma_z}(M_x)=\chi_{\sigma_z}(M_d)=-1$. The $B_{1g}$ form factor is even under $M_x$ and odd under $M_d$, $B_{2g}$ has the opposite mirror characters, and both change sign under $C_4$. The $A_{2g}$ form factor is invariant under $C_4$ and odd under both mirrors, so its product with $\sigma_z$ is preserved by every in-plane operation and both parts lie in the same $4n$ sine harmonics. This gives Table~\ref{tab:representations}.

\section*{S4. Finite-momentum current vertices}
In this section, we construct the current vertices at finite photon momentum. We couple the field through straight-bond Peierls phases. In a fixed orbital basis $H(\bm k)=\sum_{\bm\delta}h_{\bm\delta}e^{i\bm k\cdot\bm\delta}$ with $h_{-\bm\delta}=h_{\bm\delta}^\dagger$, and for charge $-e$, the Hamiltonian is
\begin{equation}
 \hat H[\bm A]=\sum_{\bm R,\bm\delta}c_{\bm R}^\dagger h_{\bm\delta}c_{\bm R+\bm\delta}
 \exp\!\left[\frac{ie}{\hbar}\int_0^1ds\,\bm\delta\cdot\bm A(\bm R+s\bm\delta,t)\right],
\label{eq:speierls}
\end{equation}
where $\bm R$ runs over lattice sites, $\bm\delta$ over bond vectors, and the path is the straight segment joining the two sites. Expanding in the field generates all vertices. The one-photon energy vertex is the continuous-path average
\begin{equation}
 \mathcal V^a(\bm k,\bm q_\parallel)=\int_{-1/2}^{1/2}ds\,\partial_{k_a}H(\bm k+s\bm q_\parallel)
 =\frac{2\sinh(D_{\bm q_\parallel}/2)}{D_{\bm q_\parallel}}\,\partial_{k_a}H(\bm k),
 \qquad D_{\bm q_\parallel}=q_{\parallel,x}\partial_{k_x}+q_{\parallel,y}\partial_{k_y},
\label{eq:sV}
\end{equation}
where $D_{\bm q_\parallel}$ is the derivative along the in-plane photon momentum. Equation~\eqref{eq:sV} is exact in $\bm q_\parallel$ and regular at $\bm q_\parallel=0$, since $2\sinh(z/2)/z=1+z^2/24+\cdots$. Equivalently, each bond carries its own weight, $\mathcal V^a=\sum_{\bm\delta}h_{\bm\delta}e^{i\bm k\cdot\bm\delta}(i\delta_a)\operatorname{sinc}(\bm q_\parallel\cdot\bm\delta/2)$, and for the nearest-neighbor axis bonds the two Cartesian components acquire different weights,
\begin{equation}
 \mathcal V^x=S_x\,\partial_{k_x}H,\qquad
 \mathcal V^y=S_y\,\partial_{k_y}H,\qquad
 S_a\equiv\operatorname{sinc}(q_{\parallel,a} a/2),
\label{eq:ssinc}
\end{equation}
where the subscript $a$ is Cartesian and the second $a$ is the lattice constant. Contracting with the in-plane photon momentum gives the exact lattice Ward identity \cite{McKay}
\begin{equation}
 q_{\parallel,a}\mathcal V^a=H(\bm k+\bm q_\parallel/2)-H(\bm k-\bm q_\parallel/2),
\label{eq:sward}
\end{equation}
which fixes the longitudinal projection, while the straight-bond Peierls coupling supplies the transverse components. The velocity vertex is $v^a=\mathcal V^a/\hbar$ and the mixed current--field vertex is $\mathcal W^{ia}=\partial_{k_i}\mathcal V^a$. Notice that the mixed vertex carries one factor of $S_a$, while the two-photon vertex $\mathcal V^{aa}(\bm k;\bm q_\parallel,-\bm q_\parallel)=S_a^2\partial_{k_a}^2H$ carries two; using a single local second derivative for both is not exact at finite $q_\parallel$.

\section*{S5. Projector algebra and the optical bilinear}
In this section, we express all matrix elements through band projectors, so that no eigenvector phase is required. We write the two-band Hamiltonian as $H(\bm k)=d_0\sigma_0+\bm d\cdot\bm\sigma$ and use symmetric endpoints $\bm k_\pm=\bm k\pm\bm q_\parallel/2$, with $\alpha=(m,\bm k_-)$ and $\beta=(n,\bm k_+)$. The band projectors are
\begin{equation}
 P_X=\tfrac12(\sigma_0+\xi_X\hat{\bm n}_X\cdot\bm\sigma),\qquad \xi_X=\pm1,
 \qquad \hat{\bm n}_\pm\equiv\hat{\bm d}(\bm k_\pm),
\label{eq:sproj}
\end{equation}
where $\xi_X$ is the band index, $\hat{\bm n}_\beta=\hat{\bm n}_+$, and $\hat{\bm n}_\alpha=\hat{\bm n}_-$, with $\operatorname{Tr}P_X=1$. These rank-one forms hold away from $\bm d=0$; at a band touching the single-band projector is undefined and the degenerate subspace must be treated as a whole. In Pauli components the vertices read
\begin{equation}
 \mathcal V^a=u_a\sigma_0+\bm w_a\cdot\bm\sigma,\qquad
 \mathcal W^{ab}=\tilde u_{ab}\sigma_0+\tilde{\bm w}_{ab}\cdot\bm\sigma,
\label{eq:svertexpauli}
\end{equation}
with $u_a=S_a\partial_{k_a}d_0$ and $\bm w_a=S_a\partial_{k_a}\bm d$. For the mixed vertex, $\tilde u_{ia}=S_a\partial_{k_i}\partial_{k_a}d_0$ and $\tilde{\bm w}_{ia}=S_a\partial_{k_i}\partial_{k_a}\bm d$, with no sum over the optical index $a$; the endpoint band velocities remain ordinary local derivatives. Using $\hat{\bm n}\cdot\partial_i\hat{\bm n}=0$, we define
\begin{equation}
 \bm G\equiv\partial_i\hat{\bm n}|_{\bm k_+},\qquad
 \bm g\equiv\partial_i\hat{\bm n}|_{\bm k_-},\qquad
 \partial_i\hat{\bm n}=\frac1d\left(\mathbb I-\hat{\bm n}\hat{\bm n}^{T}\right)\partial_i\bm d,
\label{eq:sndot}
\end{equation}
and find
\begin{equation}
\begin{aligned}
 P_\beta\partial_iP_\beta&=\tfrac{\xi_\beta}4\left[\bm G+i\xi_\beta\hat{\bm n}_+\times\bm G\right]\cdot\bm\sigma,&
 (\partial_iP_\beta)P_\beta&=\tfrac{\xi_\beta}4\left[\bm G-i\xi_\beta\hat{\bm n}_+\times\bm G\right]\cdot\bm\sigma,\\
 P_\alpha\partial_iP_\alpha&=\tfrac{\xi_\alpha}4\left[\bm g+i\xi_\alpha\hat{\bm n}_-\times\bm g\right]\cdot\bm\sigma,&
 (\partial_iP_\alpha)P_\alpha&=\tfrac{\xi_\alpha}4\left[\bm g-i\xi_\alpha\hat{\bm n}_-\times\bm g\right]\cdot\bm\sigma,
\end{aligned}
\label{eq:sPdP}
\end{equation}
all of which have vanishing scalar part. The two orderings $P\partial_iP$ and $(\partial_iP)P$ differ by the sign of the $i\,\hat{\bm n}\times\partial_i\hat{\bm n}$ term and must not be interchanged.

The optical bilinear is
\begin{equation}
 M^{jk}_{\rm raw}\equiv\operatorname{Tr}\!\left[P_\beta\mathcal V^jP_\alpha\mathcal V^k\right]
 =\hbar^2v^j_{\beta\alpha}v^k_{\alpha\beta},
\label{eq:smraw}
\end{equation}
which follows by inserting $P=|u\rangle\langle u|$. Pairing $X=P_\beta\mathcal V^j$ and $Y=P_\alpha\mathcal V^k$ and setting $p^0=2X_0$, $\bm p=2\bm X$, $r^0=2Y_0$, and $\bm r=2\bm Y$, we obtain
\begin{equation}
\begin{aligned}
 p^0&=u_j+\xi_\beta(\hat{\bm n}_+\!\cdot\bm w_j),&
 \bm p&=\bm w_j+\xi_\beta u_j\hat{\bm n}_++i\,\xi_\beta\hat{\bm n}_+\times\bm w_j,\\
 r^0&=u_k+\xi_\alpha(\hat{\bm n}_-\!\cdot\bm w_k),&
 \bm r&=\bm w_k+\xi_\alpha u_k\hat{\bm n}_-+i\,\xi_\alpha\hat{\bm n}_-\times\bm w_k,
\end{aligned}
\qquad
 M^{jk}_{\rm raw}=\tfrac12\left(p^0r^0+\bm p\cdot\bm r\right).
\label{eq:sbilinear}
\end{equation}
With $\chi\equiv\xi_\beta\xi_\alpha\,\hat{\bm n}_+\!\cdot\hat{\bm n}_-$, $A_j\equiv\xi_\beta(\hat{\bm n}_+\!\cdot\bm w_j)$, and $B_j\equiv\xi_\alpha(\hat{\bm n}_-\!\cdot\bm w_j)$, the real part is
\begin{equation}
 \operatorname{Re}M^{jk}_{\rm raw}=\tfrac12\Big[u_ju_k(1+\chi)+(\bm w_j\!\cdot\bm w_k)(1-\chi)
 +u_j(A_k+B_k)+u_k(A_j+B_j)+A_jB_k+A_kB_j\Big],
\label{eq:sReM}
\end{equation}
valid for arbitrary $\xi_\alpha$ and $\xi_\beta$. All imaginary parts follow from a single identity: for real Hermitian $A=(a_0,\bm a)$ and $B=(b_0,\bm b)$, with $[\bm A,\bm B,\bm C]\equiv\bm A\cdot(\bm B\times\bm C)$,
\begin{equation}
 \operatorname{Im}\operatorname{Tr}\left[P_\beta AP_\alpha B\right]
 =\tfrac12\Big\{[\xi_\beta\hat{\bm n}_+-\xi_\alpha\hat{\bm n}_-,\bm a,\bm b]
 -\xi_\alpha\xi_\beta a_0[\hat{\bm n}_-,\hat{\bm n}_+,\bm b]
 +\xi_\alpha\xi_\beta b_0[\hat{\bm n}_-,\hat{\bm n}_+,\bm a]\Big\}.
\label{eq:smasterim}
\end{equation}
Setting $(A,B)=(\mathcal V^j,\mathcal V^k)$ gives $\operatorname{Im}M^{jk}_{\rm raw}$, and setting $(A,B)=(\mathcal W^{ij},\mathcal V^k)$ or $(\mathcal V^j,\mathcal W^{ik})$ gives the contact terms of S7. For $j=k$, $\operatorname{Im}M^{jj}_{\rm raw}=0$ identically.

The exact transition energy at finite $q_\parallel$ is
\begin{equation}
 \Delta_{\xi_\beta\xi_\alpha}=\left(d_0^+-d_0^-\right)+\xi_\beta d_+-\xi_\alpha d_-,
\label{eq:sDelta}
\end{equation}
where $d_\pm=|\bm d(\bm k_\pm)|$ and $d_0^\pm=d_0(\bm k_\pm)$. For interband transitions it is a sum of the two gaps plus a tilt from $d_0$, while for same-band transitions it is a difference that vanishes as $\bm q_\parallel\to0$. Both are evaluated at the displaced endpoints without expanding in $q_\parallel$.

\section*{S6. Injection and shift currents}
\emph{Injection}.---The diagonal velocities carry no momentum transfer, so the Hellmann--Feynman theorem applies exactly,
\begin{equation}
 \hbar\Delta v^i=\left(u_i^+-u_i^-\right)
 +\left(\xi_\beta\hat{\bm n}_+\!\cdot\bm w_i^+-\xi_\alpha\hat{\bm n}_-\!\cdot\bm w_i^-\right),
\label{eq:sdeltav}
\end{equation}
where the first bracket is the particle--hole-asymmetric part set by $d_0$. With $f_{\beta\alpha}=f_\beta-f_\alpha$, the injection conductivity is
\begin{equation}
 \sigma^{i;jk}_{\rm inj}=\darkred{+}\frac{\pi e^3}{\hbar^2\omega^2\eta}\sum_{nm}\int_{\bm k}
 f_{\beta\alpha}\left(v^i_{\beta\beta}-v^i_{\alpha\alpha}\right)v^j_{\beta\alpha}v^k_{\alpha\beta}
 \,\delta(\omega_{\beta\alpha}-\omega)
 =\darkred{+}\frac{\pi e^3}{2\hbar^5\omega^2\eta}\sum_{nm}\int_{\bm k}
 f_{\beta\alpha}\left(\hbar\Delta v^i\right)\left(p^0r^0+\bm p\cdot\bm r\right)\delta(\omega_{\beta\alpha}-\omega),
\label{eq:sinj}
\end{equation}
where $\eta$ is the relaxation rate. The linear current takes the real part and the circular current the imaginary part of the bilinear. Since $j(t)=j\,e^{2\eta t}$ gives $dj/dt=2\eta j$, Eq.~\eqref{eq:sinj}, which already carries $1/\eta$, is the accumulated injection current rather than its rate. In the numerics we replace $\eta$ by the dc width, $\tau_{\rm inj}=\hbar/\Gamma_0$, equivalently $\tau_{\rm inj}=1/2\eta$; stating which pole is meant removes a factor-of-two ambiguity. Only the injection current carries $1/\eta$.

\emph{Shift}.---With $A\equiv v^j_{\beta\alpha}$, $B\equiv v^k_{\alpha\beta}$, and $\Delta r^i\equiv r^i_{\beta\beta}-r^i_{\alpha\alpha}$, the shift numerator is
\begin{equation}
 S^{i;jk}\equiv A\,B_{;i}-B\,A_{;i}=A\,\partial_iB-B\,\partial_iA+2i\,\Delta r^iAB,
\label{eq:sS}
\end{equation}
in which $\Delta r^i$ and $\partial_iA$ are separately gauge dependent while $S$ is not. To make this explicit, we shift both endpoints by $p\hat e_i$ at fixed \darkred{$\bm q_\parallel$} and define the loop
\begin{equation}
 \mathcal L(p)=\langle u_\beta(\bm k)|u_\beta(\bm k+p)\rangle\,v^j_{\beta\alpha}(\bm k+p)\,
 \langle u_\alpha(\bm k+p)|u_\alpha(\bm k)\rangle\,v^k_{\alpha\beta}(\bm k).
\label{eq:sloop}
\end{equation}
Every state appears once as a bra and once as a ket, so all phases cancel and $\mathcal L$ is gauge invariant. Using $\langle u|\partial_iu\rangle=-i\,r^i_{nn}$, we find $\mathcal L(p)=AB+p\,B\,A_{;i}$, hence $\partial_p\mathcal L|_0=B\,A_{;i}$; for the mirror loop $\tilde{\mathcal L}$, $\partial_p\tilde{\mathcal L}|_0=A\,B_{;i}$, and
\begin{equation}
 S^{i;jk}=\partial_p\tilde{\mathcal L}\big|_0-\partial_p\mathcal L\big|_0 .
\label{eq:sSloop}
\end{equation}
In trace form,
\begin{equation}
 \mathcal L(p)=\frac1{\hbar^2}\operatorname{Tr}\!\left[
 P_\beta(\bm k)P_\beta(\bm k+p)\,\mathcal V^j(\bm k+p)\,P_\alpha(\bm k+p)P_\alpha(\bm k)\,\mathcal V^k(\bm k)\right],
\label{eq:sloopTr}
\end{equation}
in which $p$ appears in three places; freezing $\mathcal V^j$ would omit exactly the higher-derivative vertex contained in $\partial_{k_i}v^j$. Using $P^2=P$, we obtain
\begin{equation}
\begin{aligned}
 \hbar^2\partial_p\mathcal L\big|_0&=
 \underbrace{\operatorname{Tr}\!\left[(P_\beta\partial_iP_\beta)\mathcal V^jP_\alpha\mathcal V^k\right]}_{C_1}
 +\underbrace{\operatorname{Tr}\!\left[P_\beta(\partial_i\mathcal V^j)P_\alpha\mathcal V^k\right]}_{D_j}
 +\underbrace{\operatorname{Tr}\!\left[P_\beta\mathcal V^j(\partial_iP_\alpha)P_\alpha\mathcal V^k\right]}_{C_2},\\
 \hbar^2\partial_p\tilde{\mathcal L}\big|_0&=
 \underbrace{\operatorname{Tr}\!\left[(P_\alpha\partial_iP_\alpha)\mathcal V^kP_\beta\mathcal V^j\right]}_{C_3}
 +\underbrace{\operatorname{Tr}\!\left[P_\alpha(\partial_i\mathcal V^k)P_\beta\mathcal V^j\right]}_{D_k}
 +\underbrace{\operatorname{Tr}\!\left[P_\alpha\mathcal V^k(\partial_iP_\beta)P_\beta\mathcal V^j\right]}_{C_4},
\end{aligned}
\label{eq:sC1C4}
\end{equation}
so that
\begin{equation}
 \hbar^2S^{i;jk}=C_3+C_4+D_k-C_1-C_2-D_j,
\qquad
 \sigma^{i;jk}_{\rm sh}=\darkred{-}\frac{i\pi e^3}{\hbar^4\omega^2}\sum_{nm}\int_{\bm k}
 f_{\beta\alpha}\left[C_3+C_4+D_k-C_1-C_2-D_j\right]\delta(\omega_{\beta\alpha}-\omega).
\label{eq:sshift}
\end{equation}
Because of the explicit $i$, the linear shift current takes the imaginary part of the bracket and the circular shift current the real part, opposite to the injection current. Under $\alpha\leftrightarrow\beta$ and $j\leftrightarrow k$ one has $C_1\mapsto C_3$ and $C_2\mapsto C_4$; in particular, $C_4$ is not obtained from $C_1$. For real linear polarization, Eq.~\eqref{eq:sshift} reduces to the transition rate multiplied by the gauge-invariant shift vector \cite{Resta}, which is the form ${\cal K}^a_{\rm sh}$ of Eq.~\eqref{eq:kernels}, with $|M_{fi}|^2R^a_{fi}=-\operatorname{Im}(M_{fi}^*D_aM_{fi})$ the product to use where the matrix element vanishes.

\section*{S7. Fermi-surface and contact principal-value currents}
In this section, we derive the two dispersive contributions. The second-order current is
\begin{equation}
j^{(2)}=\operatorname{Tr}\!\left[\rho^{(2)}J^{(0)}+\rho^{(1)}J^{(1)}+\rho^{(0)}J^{(2)}\right].
\label{eq:sjajb}
\end{equation}
Here $\rho^{(2)}$ is driven by both $[H^{(1)},\rho^{(1)}]$ and $[H^{(2)},\rho^{(0)}]$. For the $C_{2z}$-symmetric all-in-plane model considered here, the $\rho^{(0)}J^{(2)}$ contribution and the $H^{(2)}\rho^{(0)}$ drive vanish after Brillouin-zone integration, leaving $j_a=\operatorname{Tr}[\rho^{(2)}J^{(0)}]$ and $j_b=\operatorname{Tr}[\rho^{(1)}J^{(1)}]$ as the nonzero pieces used below. The principal part of $j_a$ gives the occupation-gradient term,
\begin{equation}
 \sigma^{i;jk}_{\rm FS}=+\frac{e^3}{\hbar^3\omega^2}\sum_{nm}\int_{\bm k}
 \left(\partial_{k_i}f_{\beta\alpha}\right)M^{jk}_{\rm raw}\,
 \mathcal P\frac1{\Delta-\hbar\omega},
\label{eq:sfs}
\end{equation}
which vanishes when the occupations are constant within completely filled or empty bands. We reserve the label FS for this term.

The contact term warrants more care. With the vertices of S5, we define
\begin{equation}
 D_j=\operatorname{Tr}\!\left[P_\beta\mathcal W^{ij}P_\alpha\mathcal V^k\right]=\hbar^3v^{ij}_{\beta\alpha}v^k_{\alpha\beta},
 \qquad
 D_k=\operatorname{Tr}\!\left[P_\alpha\mathcal W^{ik}P_\beta\mathcal V^j\right]=\hbar^3v^{ik}_{\alpha\beta}v^j_{\beta\alpha},
\label{eq:sDjDk}
\end{equation}
which are the same objects that appear in Eq.~\eqref{eq:sC1C4}. For the negative-frequency ordering one relabels both the band indices and the optical indices, giving $f_m(\bm k_-)-f_n(\bm k_+)=-f_{\beta\alpha}$ and $\epsilon_m(\bm k_-)-\epsilon_n(\bm k_+)+\hbar\omega-i0^+=-[\Delta-\hbar\omega+i0^+]$. The minus signs cancel, and the two orderings are exactly
\begin{equation}
 \frac{D_k}{\Delta-\hbar\omega-i0^+}+\frac{D_j}{\Delta-\hbar\omega+i0^+}.
\label{eq:scontactpair}
\end{equation}
Notice that the second denominator is not $\Delta+\hbar\omega$. At finite causal width $\Gamma>0$ and with $x=\Delta-\hbar\omega$, the exact identity
\begin{equation}
 \frac{D_k}{x-i\Gamma}+\frac{D_j}{x+i\Gamma}
 =(D_j+D_k)\frac{x}{x^2+\Gamma^2}+i(D_k-D_j)\frac{\Gamma}{x^2+\Gamma^2}
\label{eq:scontactsplit}
\end{equation}
separates, as $\Gamma\to0^+$, a principal-value part $(D_j+D_k)\mathcal P(\Delta-\hbar\omega)^{-1}$ from a resonant part $i\pi(D_k-D_j)\delta(\Delta-\hbar\omega)$. The resonant part contributes $-i\pi e^3\hbar^{-4}\omega^{-2}f_{\beta\alpha}(D_k-D_j)\delta(\omega_{\beta\alpha}-\omega)$, which is exactly the $D_k-D_j$ content of the shift kernel in Eq.~\eqref{eq:sshift}, and must therefore be counted once, inside the shift current. Ref.~\cite{Xie} retains this resonant part and explicitly neglects the principal-value part, which we keep,
\begin{equation}
 \sigma^{i;jk}_{b{\rm PV}}=\darkred{-}\frac{e^3}{\hbar^3\omega^2}\sum_{nm}\int_{\bm k}
 f_{\beta\alpha}\left(D_j+D_k\right)\mathcal P\frac1{\Delta-\hbar\omega}.
\label{eq:sbpv}
\end{equation}
This is the term ${\cal K}^a_{b{\rm PV}}$ of Eq.~\eqref{eq:kernels}, with $\operatorname{Re}(M_{fi}^*B^a_{fi})$ the band-projected form of $\tfrac12(D_j+D_k)$ for real linear polarization. Defining $\sigma_{\rm PV}\equiv\sigma_{\rm FS}+\sigma_{b{\rm PV}}$ and integrating the FS term by parts gives the equivalent exact form
\begin{equation}
 \sigma^{i;jk}_{\rm PV}=\darkred{-}\frac{e^3}{\hbar^3\omega^2}\sum_{nm}\int_{\bm k}f_{\beta\alpha}
 \left\{\partial_{k_i}\!\left[M^{jk}_{\rm raw}\,\mathcal P\frac1{\Delta-\hbar\omega}\right]
 +(D_j+D_k)\,\mathcal P\frac1{\Delta-\hbar\omega}\right\},
\label{eq:spvcompleted}
\end{equation}
where the derivative is understood distributionally when the resonance is accessible. The explicit contact term does not cancel the contact content generated by the derivative, since
\begin{equation}
 \partial_{k_i}M^{jk}_{\rm raw}
 =\operatorname{Tr}\!\left[(\partial_iP_\beta)\mathcal V^jP_\alpha\mathcal V^k\right]
 +\operatorname{Tr}\!\left[P_\beta\mathcal V^j(\partial_iP_\alpha)\mathcal V^k\right]
 +D_j+D_k .
\label{eq:sdMraw}
\end{equation}
We do not call $\sigma_{\rm PV}$ a Fermi-surface current: $\sigma_{\rm FS}$ carries $\partial_{k_i}f_{\beta\alpha}$ and vanishes for constant occupations, whereas $\sigma_{b{\rm PV}}$ carries $f_{\beta\alpha}$ and does not vanish for a filled valence band and an empty conduction band. The label $b{\rm PV}$ refers to this principal-value contact term and is unrelated to the use of ``PV'' for photovoltaic in Ref.~\cite{Gassner}. We treat these two dispersive terms separately due to their distinct microscopic origins; their symmetry behavior is unaffected by how the principal-value contribution is restructured.

\section*{S8. Domain parity of the microscopic terms}
In this section, we derive the domain parities used in the main text. Let $J$ enter through a time-reversal-odd term. Two relations generate the entire selection structure,
\begin{equation}
 C_{2z}H(\bm k;J)C_{2z}^{-1}=H(-\bm k;J),\qquad
 \Theta H(\bm k;J)\Theta^{-1}=H(-\bm k;-J),
\label{eq:ssymops}
\end{equation}
where $\Theta$ is time reversal. From $C_{2z}$ with three in-plane indices and $R=\operatorname{diag}(-1,-1)$,
\begin{equation}
 \sigma^{i;jk}(\bm q_\parallel;J)=R_{ii'}R_{jj'}R_{kk'}\sigma^{i'j'k'}(R^{-1}\bm q_\parallel;J)=-\sigma^{i;jk}(-\bm q_\parallel;J),
 \qquad \sigma^{i;jk}(\bm 0;J)=0,
\label{eq:sc2z}
\end{equation}
so the entire second-order response is odd in $\bm q_\parallel$ and the $q_\parallel=0$ bulk photovoltaic effect vanishes. This requires all three indices in the plane; for a mixed component such as $\sigma^{y;xz}$ the tensor factors give $(-1)(-1)(+1)=+1$, and a $q_\parallel=0$ response is allowed. Composing the two relations at fixed $J$ gives
\begin{equation}
\begin{aligned}
 \sigma_{{\rm sh},L}(-J)&=-\sigma_{{\rm sh},L}(J),&
 \sigma_{{\rm inj},L}(-J)&=+\sigma_{{\rm inj},L}(J),&
 \sigma_{{\rm PV},L}(-J)&=+\sigma_{{\rm PV},L}(J),\\
 \sigma_{{\rm sh},C}(-J)&=+\sigma_{{\rm sh},C}(J),&
 \sigma_{{\rm inj},C}(-J)&=-\sigma_{{\rm inj},C}(J),&
 \sigma_{{\rm PV},C}(-J)&=-\sigma_{{\rm PV},C}(J),
\end{aligned}
\label{eq:sparitytable}
\end{equation}
where both halves of $\sigma_{\rm PV}$ obey the same parity. The explicit factor of $i$ in Eq.~\eqref{eq:sshift} places the shift current opposite to the other terms in each polarization.

In the model of Eq.~\eqref{eq:model}, these parities have a transparent algebraic origin. In the split notation $\bm a=(\lambda\bm a_\parallel,Ja_z)$, with $\{u,v\}=u_1v_2-u_2v_1$,
\begin{equation}
 [\bm A,\bm B,\bm C]=A_z\{\bm B_\parallel,\bm C_\parallel\}+B_z\{\bm C_\parallel,\bm A_\parallel\}+C_z\{\bm A_\parallel,\bm B_\parallel\},
\label{eq:striple}
\end{equation}
and two out-of-plane components cannot contribute simultaneously. Exactly one factor therefore supplies its $J$ component and the other two supply their $\lambda$ components, so every triple product carries $\lambda^2J$, while every dot product is even in $J$. Imaginary parts of the projector traces are built from triple products and real parts from dot products; since the shift current carries an explicit $i$ and the other terms do not, Eq.~\eqref{eq:sparitytable} follows from the algebra. The $\lambda^2J$ dependence stems directly from the scalar triple product: each nonvanishing contribution contains one $z$ component proportional to $J$ and two $x$-$y$ components proportional to $\lambda$.

Two exact parameter relations complete the picture. At $\lambda=0$, $U_x=\sigma_x$ gives $U_xH(\bm k;0,J)U_x^{-1}=H(\bm k;0,-J)$ and maps every charge vertex, $U_x\mathcal V^a(0,J)U_x^{-1}=\mathcal V^a(0,-J)$, and likewise $\mathcal W^{ab}$. Since $U_x$ is momentum independent, the endpoint projectors and their derivatives transform in the same way, and unitary invariance of the trace gives $\sigma_{\rm charge}(0,J)=\sigma_{\rm charge}(0,-J)$. Every domain-odd charge current therefore vanishes without SOC. Separately, $U_z=\sigma_z$ gives $\sigma_{\rm charge}(\lambda,J)=\sigma_{\rm charge}(-\lambda,J)$. This result is specific to charge currents: for $s^z\propto\sigma_z$ one has $U_xs^zU_x^{-1}=-s^z$, so no $\lambda=0$ null follows for a spin current. Neither relation establishes analyticity at $\lambda=0$, and we do not assert a small-$\lambda$ scaling law.

\end{document}